# Graphene stabilization of two-dimensional gallium nitride

Zakaria Y. Al Balushi[1,2], Ke Wang[3], Ram Krishna Ghosh[2,4], Rafael A. Vilá[1,2], Sarah M. Eichfeld[1,2,3], Joshua D. Caldwell[5], Xiaoye Qin[6], Yu-Chuan Lin[1,2], Paul A. DeSario[5], Shruti Subramanian[1,2], Dennis F. Paul[7], Robert M. Wallace[6], Suman Datta[2,3,4], Joan M. Redwing[1,2,3,4,a] and Joshua A. Robinson[1,2,3,b]

[1]Department of Materials Science and Engineering, The Pennsylvania State University, University Park, PA, 16802 USA
[2]Center for 2-Dimensional and Layered Materials, The Pennsylvania State University, University Park, PA, 16802 USA
[3]Materials Research Institute, The Pennsylvania State University, University Park, PA, 16802 USA
[4]Department of Electrical Engineering, The Pennsylvania State University, University Park, PA, 16802 USA
[5]U.S. Naval Research Laboratory, Washington, D.C., 20375 USA
[6]Department of Materials Science and Engineering, The University of Texas at Dallas, Richardson, TX, 75080 USA
[7]Physical Electronics USA, 18725 Lake Drive East, Chanhassen, MN, 55317 USA

Corresponding authors: [a] e-mail: jmr31@psu.edu, [b] e-mail: jrobinson@psu.edu

**The spectrum of two-dimensional (2D) materials "beyond graphene" offers a remarkable platform to study new phenomena in condensed matter physics. Among these materials, layered hexagonal boron nitride (hBN), with its wide bandgap energy (~5.0-6.0 eV), has clearly established that 2D nitrides are key to advancing novel devices[1]. A gap, however, remains between the theoretical prediction of 2D nitrides "beyond hBN"[2,3] and experimental realization of such structures. Here we demonstrate the synthesis of 2D gallium nitride (GaN) *via* a novel migration-enhanced encapsulated growth (MEEG) technique utilizing epitaxial graphene. We theoretically predict and experimentally validate that the atomic structure of 2D GaN grown *via* MEEG is notably different from reported theory[2–4]. Moreover, we establish that graphene plays a critical role in stabilizing the direct-bandgap (nearly 5.0 eV), 2D buckled structure. Our results provide a foundation for discovery and stabilization of novel 2D nitrides that are difficult to prepare *via* traditional synthesis.**

Group-III nitride devices have been explored for many years, yet in order to realize novel applications such as tunnel junctions[5], single photon emitters[6,7] and polarization-driven topological insulators[8], the growth of stoichiometric atomic layers of group-III nitrides must be explored. Interestingly, wurtzite GaN is predicted to reconstruct into a 2D hexagonal graphitic structure when thinned to a few atomic layers[2–4,9], which leads to a thickness-dependent energy bandgap ($E_g$) *via* quantum confinement[10]. The experimental realization of large-area 2D nitrides "beyond hBN" on technologically relevant substrates, however, has not come to fruition[11]. This is because obtaining atomic layers by cleaving tetrahedral coordinated bulk crystals is quite challenging, and results in unsaturated dangling bonds on the surface[12]. To satisfy the surface electrostatics, these cleaved layers take on trigonal planar coordinated graphitic structures by means of charge compensation through surface reconstruction, electron redistribution and/or adsorption of charged species[13]. Though charge compensation in cleaved atomic layers of group-III nitrides may be reached through electron redistribution, the resulting surface energy, in most cases, is high. Therefore, new pathways to efficiently stabilize graphitic structures of group-III nitrides should be considered.

The stability of cleaved wurtzite surfaces is affected by surface passivation. Two types of freestanding monolayer hexagonal group-III nitride structures, planar and buckled (Fig. 1a), are predicted to be stable. Based on density functional theory (DFT, see Methods), when unsaturated states are not properly passivated, a planar structure is obtained; however, when two layers of planar 2D nitrides come into contact, bonding beyond van der Waals (vdW) occurs, leading to a semi-metallic behavior as reflected in the density of states (DOS, Supplementary Fig. 1). Alternatively, when unsaturated states are properly passivated (using partially charged pseudo-hydrogen)[14], the structure is most stable in the buckled form. To identify the most stable structure, the binding energy of the 2D planar and buckled structures as a function of layer number was investigated. Evident from Figure 1a, the freestanding buckled structure exhibits a more negative binding energy when compared to the planar structure, and therefore is considered the preferred configuration for 2D nitrides, regardless of the number of layers. Such variations in atomic configuration lead to significant differences in the electronic structure. Perhaps most immediately notable in freestanding monolayer 2D GaN is the prominent increase in $E_g$ compared to bulk GaN (3.42 eV) due to quantum confinement[10]. More importantly, we find that the planar structure (Fig. 1b) exhibits an indirect $E_g$ of 4.12 eV, while the buckled structure maintains a direct $E_g$ of 5.28 eV. This is a critical difference between the predicted 2D GaN layers (direct *versus* indirect $E_g$), as it has significant impact on the realization of optoelectronic devices. These results demonstrate that the direct $E_g$ buckled structure is the most thermodynamically stable, indicating that 2D nitrides (and alloys) are viable candidates for tunable optoelectronics (Fig. 1c).

Direct growth of wurtzite GaN on SiC(0001) conventionally results in three-dimensional (3D) island formation due surface energy constraints and large lattice mismatch. If one, however, is able to passivate sites of high surface energy to promote Frank–van der Merwe growth, 2D GaN can be realized. Here, we introduce a novel "Migration-Enhanced Encapsulated Growth" (MEEG) process to enable the growth of 2D GaN (see Methods). A schematic of the proposed MEEG process is shown in Figure 2a-c. The starting substrate consists of epitaxial graphene formed by sublimation of silicon from the surface of SiC(0001) (Fig. 2a), which is converted to quasi-freestanding epitaxial graphene (QFEG) *via* hydrogenation[15]. Hydrogenation passivates dangling bonds between graphene-buffer/SiC(0001), converting the graphene-buffer to an additional layer of graphene (Figure 2b)[16] and thus creating a pristine interface of reduced energy for the realization of 2D nitrides *via* MEEG. QFEG/Si(0001) is then exposed to cycles of trimethylgallium at 550°C (Fig. 2c), which decomposes to gallium adatoms that diffuse readily on the surface of graphene, subsequently intercalating between QFEG/SiC(0001). Finally, transformation of the intercalated gallium to 2D GaN is performed *via* ammonolysis at 675°C. During this process, atomic nitrogen resulting from the decomposition of ammonia, intercalates graphene and reacts with gallium to form 2D GaN[17]. Ammonolysis of gallium on SiC(0001) *without* a graphene capping layer results in the formation of 3D doughnut-shaped structures of GaN due to enhanced desorption of gallium from the apex of liquid gallium droplets prior to ammonolysis (Supplementary Fig. 2a). In contrast, when QFEG is utilized, GaN wets the surface in a patchwork of regions in addition to forming 3D structures on the surface of graphene (Supplementary Fig. 2b). The regions of bright contrast in scanning electron micrographs are found to consist of gallium and nitrogen when evaluated by Auger electron spectroscopy (AES, Supplementary Fig. 2c-f) and exhibit a unique Raman longitudinal optical phonon mode from that of bulk GaN (Supplementary Fig. 3a). Aberration-corrected scanning transmission electron microscopy (STEM, Fig. 2d) subsequently reveals that these regions are 2D GaN, consisting of two sub-layers of gallium, located at the graphene/SiC(0001) interface. Further inspection with elemental mapping *via* energy dispersive x-ray spectroscopy of silicon (Fig. 2e), gallium (Fig. 2f), and nitrogen (Fig. 2g) in the STEM cross-section specimens confirms that these layers are GaN, and provides direct evidence that MEEG is a viable route for 2D nitride synthesis.

Defects in graphene facilitate intercalation. Similar to lithium, europium and cesium intercalation in epitaxial graphene[18–20], pre-existing point defects, wrinkles, and metal-graphene interactions serves as penetration sites during MEEG for gallium intercalation. As evidence of this, we find that the regions of 2D GaN are concentrated near 3D islands of GaN formed *via* ammonolysis of accumulated gallium (Fig. 3a) and networks of graphene wrinkles (Fig. 3b). The accumulation of gallium at defects in graphene, where nucleation of gallium nitride is preferential[21], likely provides the necessary energy to overcome the vdW interfacial binding and drive intercalation (Fig. 3c)[20,22]. Additionally, the pronounced distribution of 2D GaN preferentially near dense networks of graphene wrinkles (Supplementary Fig. 2g-h) provides evidence that gallium intercalates through nanoscale tears (Fig. 3d) that result from the large deformation at the apex[23]. Once intercalation is initiated, the passivated surface of SiC(0001) and the low migration energy of gallium on the surface of graphene (0.05 eV)[24], provide a mechanism for large diffusion across the QFEG/SiC(0001) interface. As evident in Figure 2d, graphene remains after MEEG; however, it is defective. We find a large Raman blue-shift in graphene's defect (D)-peak position relative to non-intercalated regions of graphene (~1351.9 cm$^{-1}$ to ~1331.2 cm$^{-1}$). This indicates structural changes to the graphene and possible interactions at the graphene/2D GaN interface (Supplementary Fig. 3b-d).

Two-dimensional GaN grown *via* MEEG is isostructural to layered In$_2$Se$_3$ (space group: R3m)[25], and exhibits covalent bonding to the SiC(0001) substrate while preserving a vdW gap with graphene. This is evident in annular bright field (ABF) STEM (Fig. 3e-f), which directly resolves the atomic columns[26]. Heavy atoms (gallium and silicon) in ABF-STEM appear more intensely black than light (nitrogen and carbon) atoms. As a result of the passivated QFEG/(0001)SiC interface and the pronounced spontaneous polarization of the 6H-SiC crystal, we hypothesize an initial stabilization of bilayer gallium at this interface prior to ammonolysis. This phenomenon is similar to the stabilization of bilayer gallium on the pseudo-1×1 reconstructed GaN(0001) surface in ultra-high vacuum after molecular-beam epitaxy[27], which theoretically becomes even more energetically stable when interfaced with graphene[28], supporting the observed 2D GaN structure (Fig. 3e-f) formed after ammonolysis.

In addition to 2D GaN, MEEG with QFEG/SiC(0001) enables the formation of thicker (>5 nm) layers of GaN at the interface. Interestingly, even this "thick" GaN (Fig. 4a) exhibits the same surface termination as 2D GaN observed in Figure 3e-f (Fig. 4b), thus confirming graphene's significant role in stabilizing the 2D buckled quintuple R3m structure regardless of the underlying passivated surface. The role of graphene in the formation of 2D GaN is further accentuated upon comparing ABF-STEM cross-section images of 2D GaN (Fig. 4b) to the surface of bulk GaN (Fig. 4c) grown directly on SiC(0001), where only native oxide reconstruction is observed on the surface of bulk GaN. Moreover, we find that the polarity of the nitrogen-gallium termination in 2D

GaN is inverted at the top interface compared to thick GaN (Fig. 4b). We hypothesize that during ammonolysis, nitrogen replaces hydrogen on the SiC surface, effectively passivating sites of high surface energy[29], which leads to charge neutrality and stability of 2D GaN. This is further evident upon removal of graphene and exposure of 2D GaN to air, where no detectable oxidation or change in the chemical state of 2D GaN is observed in x-ray photoelectron spectroscopy (XPS, Supplementary Fig. 4a-h).

The atomic arrangement of 2D GaN considerably impacts the electronic structure. DOS calculations of the graphene/2D GaN/SiC(0001) heterostructure (Fig. 5a), using DFT functional *meta*-generalized gradient approximation (*meta*-GGA) and Heyd-Scuseria-Ernzerhof density functionals (HSE06, see Methods), both show p-type semiconducting behavior for 2D GaN with theoretical $E_g$ values of 4.79 eV and 4.89 eV, respectively (Fig. 5b). From ultraviolet-visible reflectance measurements of 2D GaN (UV-Vis, see Methods), an inflection point in the spectra of 2D GaN at 4.90 eV is clearly evident (Fig. 5c). This specular feature is not observed in the QFEG/SiC reflectance and therefore indicates experimental observation of the $E_g$ of 2D GaN, well within the range predicted by DFT. Furthermore, we extract the absorption coefficient ($\alpha$) of 2D GaN (Fig. 5d), by fitting the dielectric function of the heterostructure collected with UV-Vis spectroscopic ellipsometry (see Methods). Using a Gaussian fit, we find a direct transition consistent with the predicted direct $E_g$ of 2D GaN centered at 4.98 eV ± 0.13 eV. In addition, low-loss electron energy-loss spectroscopy (EELS) measurements, after removal of graphene (see Methods), provides further evidence of the increased $E_g$ in 2D GaN (Supplementary Fig. 5). From low-loss EELS, the extracted $E_g$ of 2D GaN is 5.53 eV. The ~0.6 eV deviation in the $E_g$ measured from low-loss EELS arises from specular losses, such as Cerenkov loss (see Supplementary Note b), that are observed in EELS measurements of high dielectric materials. Our results, however, provide direct evidence that the $E_g$ is much larger than that found in bulk GaN, with theoretical and experimental values being in good agreement. We also carried out vertical transport measurements of the heterostructure with conductive atomic force microscopy (C-AFM) using a conductive (n-type) 6H-SiC substrate (see Methods). As shown in the current-voltage (I-V) curve in Figure 5e, under forward bias, conduction is likely from electrons being pulled from the accumulated n-type 6H-SiC over the conduction band offset between SiC and 2D GaN. From the DOS, the conduction band offset is ~1.7 V. Therefore, the turn-on voltage is within that range. Under reverse bias, conduction is likely due to electron injection from graphene over the Schottky barrier provided by 2D GaN, and into the depleted n-type 6H-SiC. In all other cases, where the 2D GaN is absent, the I-V curves exhibit an ohmic behavior (Fig.5e, inset), confirming that ambipolar graphene makes a low Schottky barrier contact with n-type 6H-SiC. Finally, the experimental observation of buckling in 2D GaN from ABF-STEM (Fig 3e-f and 4b), validates our theoretical predictions that structural buckling leads to the most thermodynamically stable structure for 2D nitrides, which retains a direct $E_g$, unlike reported theory[2–4].

Graphene has proven to be a remarkable material over the last decade; and with the discovery that it can stabilize 2D forms of traditionally "3D" binary compounds, we have provided the foundation to realize many other classes of materials that are not traditionally 2D. Furthermore, the MEEG process could enable new vertically stacked 2D layered heterostructures with yet to be predicted properties. Moreover, modifications to the phonon DOS of 2D GaN with respect to bulk GaN are anticipated to modify polaritonic behavior for additional nanophotonic functionality in the mid-infrared[30]. Finally, recognizing the impact of 2D nitrides, it can be expected that the addition of 2D GaN and other 2D nitrides (and alloys) will open up new avenues of research in novel electronic and optoelectronic devices, composed of 2D atomic layers of group-III nitride semiconductors.

## Methods

***Ab initio* density functional theory simulations**:

The calculations of binding energy and different electronic properties for 2D nitrides (GaN, AlN and InN) are performed using *ab initio* DFT as implemented in the Atomistic-ToolKit (ATK) version 2014.3 QuantumWise A/S (www.quantumwise.com). In our calculations, we optimize different structural forms by the Broyden–Fletcher–Goldfarb–Shannon (BFGS) scheme until all the forces acting on atoms are mitigated to less than 0.01 eV/Å and the stress is less than 0.01 eV/Å$^3$. A well conserved Monkhorst-Pack grid (21×21×21) is used with a mesh cutoff energy of 10 Hartree for the bulk cases whereas we use a sampling of 21×21×1 for the 2D structures. To make isolated 2D nitride layers from the cleaved [0001] bulk structure, we used a supercell having a thick vacuum region (15 Å) along the *c*-axis (while retaining the same lattice parameters '*a*' and '*b*' as in the bulk). Moreover, the vdW force between the different layers also has a significant effect in determining the interlayer distance for GaN. To incorporate the van der Waals interactions, we add a semi-empirical dispersion potential term to the conventional Kohn-Sham DFT energy through the Grimme's DFT-D2 method for all exchange correlation energies. We use generalized gradient approximation (GGA) functional in our DFT scheme to calculate the binding energies, whereas *meta*-GGA functional for the $E_g$ corrections (Supplementary Reference 1). Furthermore, hybrid functional HSE06 DOS calculations were also performed under FHI-aims function. It is computationally heavy for periodic systems. Therefore, we used a low k-point grid of 6×6×1. All other numerical parameters were the same as the *meta*-GGA functional approach.

**Synthesis of quasi-free standing epitaxial graphene and measurements**:

Both semi-insulating (vanadium doped) on-axis silicon-face 6H-SiC (*II-VI Incorporated*) and n-type (nitrogen doped) on-axis silicon-face 6H-SiC (*Cree Incorporated*) were cleaned by sonication in acetone and isopropanol alcohol followed by a heated bath in a piranha solution

(Nanostrip) and then subjected to an oxygen plasma etch (*Plasma-Therm* 720: 150 W, 10 mTorr, 45 sccm oxygen) for 2 minutes. Graphene growth conditions were adjusted in order to improve the uniformity of graphene on respective substrates (semi-insulating and n-type 6H-SiC). Substrates were annealed at 1400-1500°C (200 or 700 Torr, 5-10% $H_2$, 90-95% argon, respectively) for 30 minutes in order to obtain atomically flat terraces. The growth chamber was then evacuated and allowed to dwell at ultra-high vacuum (< $1\times10^{-9}$ Torr) for 10 minutes. The growth of epitaxial graphene commenced on the silicon-face of SiC(0001) *via* sublimation of silicon in 100% argon at 1575°C or 1625°C under 700 or 200 Torr total pressure, respectively. In the addition to the graphene-buffer layer, the resulting graphene is primarily monolayer along the (0001) terraces with some bi- and tri-layer graphene near the $(1\bar{1}0n)$ step edges of 6H-SiC. Growth quality and wrinkle density of graphene were examined using Raman spectroscopy and atomic force microscopy (AFM), respectively. Raman spectra were collected using a *WITec* Confocal Raman system operated at room temperature using a 488 nm laser excitation source. Atomic force microscopy (AFM) of graphene was performed in tapping mode using a *Bruker Dimension Icon* system (probe resonance frequency ~320 kHz) with a scan rate of 1.00 Hz. Hydrogenation of graphene was then performed for 30 minutes at 900-1100°C in 9.20 slm total flow of ultrahigh purity hydrogen under 100 Torr total pressure. The graphene-buffer layer is converted to an additional layer of graphene (QFEG) resulting in primarily bilayer graphene along the (0001) terraces of 6H-SiC.

**Synthesis of 2D GaN *via* migration-enhanced encapsulated growth (MEEG)**:

The growth of 2D GaN was realized in a customized vertical two-flow (group-III and group-V) showerhead cold-wall reactor on a graphite susceptor heated through induction. After *in situ* hydrogenation of graphene to form QFEG, samples were held at 550°C in 9.2 slm total flow of ultrahigh purity hydrogen under 50 Torr total pressure. The intercalation of gallium commenced with 60 cycles of 7.93 µmol/minute of trimethylgallium (TMGa). Each cycle consisted of a 2 second pulse of TMGa and a 3 second purge in hydrogen. Following gallium intercalation, samples were ramped to 675°C and held for 5 minutes. Ammonolysis of intercalated gallium was performed for 15 min *via* decomposition of ultrahigh purity ammonia (62.5 mmol/min) at 675°C, also in ultrahigh purity hydrogen and 50 Torr total pressure. Samples were then cooled naturally to room temperature in ultrahigh purity nitrogen.

**Post synthesis characterization of the as-grown 2D GaN samples:**

The patchwork regions of 2D GaN under graphene were investigated using a *Carl Zeiss Merlin* field emission scanning electron microscope (SEM) operated at 1.0 kV. In addition, regions of intercalated gallium and nitrogen were analyzed by Auger electron spectroscopy (AES) in the *PHI 720 Scanning Auger Nanoprobe*. SEM images with FOV of 1.0 µm were collected under 25.0 kV and 1.0 nA. Elemental maps (128×128 pixel) were then collected in AES, operated at 25.0 kV, 5.0 nA current with the sample normal to the electron gun. We compared the growth of 2D GaN *via* MEEG to bulk GaN grown directly on 6H-SiC. The conditions for the growth of bulk GaN directly on 6H-SiC are outlined in Supplementary Reference 2. Moreover, Raman spectra and mapping of the graphene/2D GaN/6H-SiC heterostructure was collected with a *Horiba LabRAM HR Evolution* confocal Raman system operated at room-temperature using (633 nm "max at 16 mW", 532 nm "max at 48 mW", 364 nm "max at 50 mW") excitation sources. For electrical measurements, 2D GaN was grown *via* MEEG using epitaxial graphene grown on n-type 6H-SiC. Site specific regions of the graphene/2D GaN/6H-SiC heterostructure were isolated using focus ion beam (FIB), in order to probe consistent vertical transport measurements without shorting *via* graphene. Measurements were performed under PeakForce TUNA mode in a *BRUKER Dimension* conductive AFM (C-AFM) using platinum AFM tip. Current-voltage (I-V) characteristics of the heterostructure were investigated by sweeping the voltage from the conductive tip between -2.5 V to 2.5 V, under a loading force and sensitivity of 5 nN and 1 nA/V, respectively.

For bandgap measurements, reflectance of samples were collected with UV-visible spectroscopy using a 60 mm integrating sphere on a *PerkinElmer LAMBDA 1050 UV/Vis/NIR* spectrophotometer (resolution of spectral bandwidth ≤ 0.05 nm) operated at room temperature. UV-Vis ellipsometeric measurements were performed using a *J.A. Woolam Co., Inc. V-VASE* spectroscopic *ellipsometer*. Data was collected over the spectral range from 1.55 eV to 6.42 eV at 40, 50, 60 and 70 degrees (see Supplementary Note a).

**High resolution electron microscopy specimen preparation, imaging and spectroscopy**:

Cross-section transmission electron microscope (TEM) specimens of regions of 2D GaN were prepared by *in situ* lift out *via* milling in *a FEI Helios NanoLab DualBeam 660* FIB. Prior to milling, thick protective amorphous carbon was deposited over the region of interest by electron beam deposition. The FIB-TEM membrane was fabricated with a starting milling voltage of 30 kV and then stepped down to 2 kV to minimize sidewall damage and thin the specimen to electron transparency. The thickness of the samples was < 30 nm, as measured by EELS. During dual-EELS measurements, a source electron monochromator was used to improve the signal-to-noise and reduce spectral delocalization at 80 kV. The energy spread full width at half maximum (FWHM) of the monochromator was setup to be < 0.2 eV. Prior to milling of specimens for $E_g$ measurements of 2D GaN *via* low loss EELS, samples were first prepared for TEM after MEEG by mechanical exfoliation of graphene layers from the surface, followed by the deposition of 30 nm of $SiO_2$ using thermal evaporation *Kurt J. Lesker Lab 18* (see Supplementary Note b).

Moreover, high resolution electron microscopy of regions of 2D GaN (STEM at 300 kV in Fig. 2d and monochromated TEM at 80 kV in Fig. 4a) was performed in a *FEI* dual aberration corrected *Titan$^3$* G2 60-300 S/TEM equipped with a SuperX energy dispersive x-ray (EDX) spectrometry system. STEM conditions were 50 pA for beam current, C2 aperture of 70 μm and camera length of 115 mm. Elemental mapping with EDX in STEM mode was performed at 300 kV with acquisition times of up to 5 minutes. The spatial resolution of aberration corrected STEM in annular bright field (ABF) mode makes it possible to directly image the position of nitrogen and gallium in the 2D GaN R3m structure. In the wurtzite structure of bulk GaN, the displacement of projected atomic pairs, referred to as dumbbells, can be resolved along the [11$\bar{2}$0] zone axis. The local polarity of the dumbbells indicates the structural nature of 2D GaN (planar or buckled). If the 2D structure is buckled, the polarity of the dumbbell, given by the terminating (gallium or nitrogen) atom facing SiC(0001) can be directly imaged. As a result, we captured simultaneous high angle annular dark field (HAADF) and ABF images at 200kV. The HADDF detector (*Fischione*) had a collection angle of 51-300 mrad for Z-contrast imaging, while the ABF detector (*FEI DF4*) had a collection angle of 7-40 mrad for atomic column displacement imaging (nitrogen and gallium dumbbells). Microscope conditions for ABF-STEM were 50 pA for beam current, C2 aperture of 70 µm and camera length of 145 mm. Optimal contrast conditions to image the nitrogen atoms in the ABF images was achieved using a negative defocus of approximately -100 Å. Finally, fast Fourier transform (FFT) bandpass filtering and Gaussian blur function were applied to select ABF-STEM images to improve visualization of atomic columns of lighter elements. Features of small and large structures were filtered up to 6 and 60 pixels, respectively and with a Gaussian blur radius of up to 2 pixels.

**Stability of 2D GaN analyzed with x-ray photoelectron spectroscopy (XPS):**

The graphene/2D GaN/semi-insulating 6H-SiC sample was loaded into an ultra-high vacuum (UHV) system which consists of multiple UHV chambers. The UHV system includes a remote plasma source in a plasma-enhance atomic layer deposition (PEALD) chamber (Picosun PR200) and an x-ray photoelectron spectroscopy (XPS) chamber. These chambers are interconnected with a UHV transfer tube all maintained at $< 3.75 \times 10^{-10}$ Torr, as described elsewhere (Supplementary Reference 3). For stability studies of 2D GaN in vacuum and air, graphene was first etched off the sample with *in situ* remote plasma, without any damage or modifications to the 2D GaN surface as confirmed by angled resolved XPS (Supplementary Reference 4). A forming gas (FG) plasma (90% $N_2$ and 10% $H_2$) at flow of 150 sccm was ignited at 2000 W for 120 seconds with the substrate temperature maintained at 200°C (Supplementary Reference 5). The XPS measurements were carried out during all steps (as-grown, after graphene FG plasma etch and after exposure to air) in order to monitor the stability of 2D GaN.

The XPS system was equipped with a monochromated Al kα (hυ = 1486.7 eV) x-ray source and a 7 channel analyzer with a pass energy of 15 eV. All high resolution XPS measurements were collected at 45° with respect to the sample normal and at same sample position. High resolution core level spectra were collected for the **C 1s**, **Si 2p**, **O 1s**, **N 1s**, **Ga 2p$_{3/2}$** and **Ga 3d** regions with detailed peak deconvolution and fitting carried out with the AAnalyer software after appropriate charge correction to compensate any shifts in the peak core level positions.

**Acknowledgements**

Materials and experimental methods in this work were partially supported by Asahi Glass Co., Ltd Japan and the National Science Foundation under grant numbers DMR-1006763 (J.M.R.), DMR-1410765 (J.M.R.), DMR-1420620 (J.M.R & J.A.R Seed program through Penn State MRSEC - Center for Nanoscale Science) and DMR-1453924 (J.A.R.). Any opinions, findings, and conclusions or recommendations expressed in this material are those of the author(s) and do not necessarily reflect the views of the National Science Foundation. Material characterization described in this work was supported by The Pennsylvania State University (PSU) Materials Characterization Laboratory (MCL) Staff Innovation Funding (SIF) program, and the Alfred P. Sloan Foundation, USA. Theoretical work (S.D.) and XPS analysis (R.M.W.) was supported by the Center for Low Energy Systems Technology (LEAST). LEAST is one of six Semiconductor Research Corporation STARnet centers sponsored by MARCO and DARPA.

We would like acknowledge Dr. Vincent Bojan (staff scientist, MCL) and Joshua Maier (technical staff, MCL) for contributions in AES analysis and FIB-TEM specimen preparation, respectively. In addition, we also acknowledge the microscopy and spectroscopy assistance provided by: Dr. Nasim Alem (PSU); Dr. Kabius Bernd (PSU); Amin Azizi (PSU); Dr. Greg Stone (PSU); Dr. Joseph Tischler (NRL); Dr. Chase Ellis (NRL); Dr. Jeffrey Owrutsky (NRL) and Dr. Takahira Miyagi (Asahi glass, Japan). Finally, we thank Dr. Tom Tiwald (J.A. Woolam) for his

## Author contributions

Experiments were designed by Z.Y.A., J.M.R. and J.A.R. The MEEG process development, SEM, Raman and data analysis (electrical, microscopic and spectroscopic) were performed by Z.Y.A. High resolution electron microscopy was conducted by K.W. Moreover, R.K.G. and S.D. designed and implemented theoretical calculations and structure simulations. R.A.V., S.M.E. and S.S. carried out epitaxial graphene growth and quality assessment. Moreover, UV-Vis measurements were performed at NRL with P.A.D. and J.D.C. J.D.C. also performed ellipsometric measurements and model development. X.Q. and R.M.W. performed the plasma processing, XPS measurements and analysis at UT Dallas. Y.-C.L. performed vertical transport measurements using C-AFM and S.D. provided input on I-V characteristics. Finally, D.F.P. performed AES measurements at PHI. All authors discussed results at all stages. Z.Y.A., J.A.R. and J.M.R. wrote the paper.

Correspondence and requests for materials should be addressed to J.M.R.[a] and/or J.A.R.[b] Corresponding authors: [a] e-mail: jmr31@psu.edu, [b] e-mail: jrobinson@psu.edu

## Competing financial interests

The authors declare no competing financial interests

**Supplement Notes:**

a) Bandgap extraction of 2D GaN via UV-vis spectroscopic ellipsometry measurements:

The dielectric function was defined by first taking the measured ellipsometric data from the control sample consisting of epitaxial graphene on a 6H-SiC substrate. The model consisted of Lorentz oscillators to define the optical response of the SiC and a 5.5 Å thick layer using the dielectric function reported by Boosalis et al. for graphene (Supplementary Reference 6). Following the establishment of a high quality fit, this dielectric function was used as the base model for the measured sample featuring the graphene/2D GaN/6H-SiC *heterostructure*. This was modeled as a single Gaussian oscillator, with the amplitude, center frequency and line width provided as variable parameters. The thickness was initially defined as 6 Å, then after initial fitting and with the Gaussian parameters also treated as a variable, a final thickness of 5.6 Å was determined. This thickness is in good agreement with STEM measurements of 2D GaN. From this fit, the center frequency was determined to be 4.98 eV. Due to the discontinuous nature of the 2D GaN, an exact dielectric function for the 2D GaN layer could not be established. However, in the context of this work, these measurements served to provide additional spectroscopic evidence of the ~4.9 eV direct bandgap transition associated with the 2D GaN as predicted by DFT.

b) Bandgap extraction of 2D GaN via low-loss EELS measurements:

Exfoliation of graphene from the surface of the samples containing 2D GaN was performed in order to minimize the potential influence of the Plasmon peaks of graphene on the EELS spectra. Dual EELS measurements in STEM mode allows collection of both high intensity low energy loss spectra (*shifted from the zero loss peak*) and weak intensity high energy loss spectra (*containing the zero loss peak*), simultaneously from the same sample position and under identical microscope conditions (Supplementary Reference 7).

This allows for the correction for any drift in the spectra by calibrating the zero loss peak energy at 0.0 eV. To extract the $E_g$ from low loss EELS, the zero loss peak and background were subtracted from the spectrum by stitching both the high energy loss and low energy loss spectrum and using a Power-law fit. No plural scattering deconvolution of our spectra was needed due the thickness of the FIB cross-section prepared sample being < 30 nm, as measured by EELS. Following careful background subtraction, a Savitzky-Golay smoothing function (second-order polynomial) was applied to the spectra. The low loss EELS spectrum relates to the joint density of states in the material. Therefore, the $E_g$ transition can be extracted from the first prominent rise in the spectrum. We used the parabolic fitting method of the first rise to extract the $E_g$ of 2D GaN, where the parabola intersects the energy axis (Supplementary Reference 8). From low-loss EELS (Supplementary Fig. 3), the extracted $E_g$ of 2D GaN is ~5.5 eV, while the Eg of $SiO_2$ is ~9.6 eV. Difference in the $E_g$ obtained *via* low-loss EELS from that collected experimentally (UV-vis reflectance and spectroscopic ellipsometry) and/or calculated theoretically (*meta*-GGA and HSE06) arises from Cerenkov loss (Supplementary Reference 7). We mitigate these effects on the spectra by acquiring the low-loss EELS spectra at 80 kV and using monochromator with an energy resolution of < 0.2 eV.

c) Stability of 2D GaN monitored by x-ray photoelectron spectroscopy (XPS):

To evaluate the stability of 2D GaN after removal of graphene, a time dependent air exposure study of the GaN chemical states was performed using XPS. XPS probes areas ~300 µm in diameter. Though mechanical exfoliation of graphene, *via* scotch tape, can remove regions of graphene sufficient for FIB prepared cross-sectional specimens used for spectroscopic measurements (Supplementary Fig. 5), *in situ* complete removal of graphene is necessary for the accurate assessment of the stability of 2D GaN by XPS, before and after removal of graphene and after exposure to air. Therefore, we have developed a process to remove graphene using a remote forming gas (FG) plasma (see

Methods) in a multi-chamber UHV system at UT Dallas (X.Q. and R.M.W., see Authors Contributions).

Complete removal of graphene was achieved after 120 second exposure to the remote FG plasma as shown in Supplementary Figure 4a. The sp$^2$ C-C bonding of graphene at 284.3 eV was near the XPS detection limit after the FG exposure, indicating a complete removal of the graphene capping layer. From prior surface treatment studies of AlGaN and GaN with remote FG plasma, removal of carbon from the surface can be achieved without any detectable chemical changes or destruction of the AlGaN and GaN surfaces (Supplementary Reference 4 and 9). Before removal of graphene, however, two distinct chemical states of gallium was detected from collected high resolution XPS spectra (labeled: Post MEEG, Supplementary Fig. 4). The higher binding energy (BE) peak corresponds to $Ga_2O_3$ (from the oxidized 3D GaN islands on the surface of graphene, Fig. 3a and b) and the lower BE peak is attributed to the underlying 2D GaN (Fig. 2d), respectively shown in Supplementary Figure 4f and g. Oxidation of such 3D islands that form on graphene was also observed in our previous report on the growth of GaN directly epitaxial graphene (Al Balushi, Z.Y. et al., Reference 20). These 3D islands, however, are easily detached from the surface by exfoliation with scotch tape and completely removed after *in situ* FG plasma selective etching of graphene, as verified in the "Post FG plasma" of the **C 1s** core level spectrum (Supplementary Fig. 4a) as well as in the associated **Ga 2p$_{3/2}$** and **Ga 3d** core level spectra (Supplementary Fig. 4f and g). After exposure to the remote FG plasma, 2D GaN is detected with the concomitant increase of the associated **Ga 2p$_{3/2}$** and **Ga 3d** peak intensities due to the removal of graphene, while the $Ga_2O_3$ concentration is near the limit of detection. Due to the removal of the graphene layer, the intensity of Si-O bonding that originates from the native oxide on SiC also increases (Supplementary Fig. 4b). Furthermore, we find exposed areas of the 6H-SiC substrate, which are not encapsulated with QFEG or graphene/2D GaN, to be comprised of Si-N bonding in the **Si 2p** core level spectrum after the remote FG plasma exposure (Supplementary Fig. 4b). Moreover, no changes in the shape or shifts in binding energy of the **O 1s** core level peak are observed for the case of graphene/2D GaN/SiC and QFEG/SiC samples after exposure to the remote FG plasma (Supplementary Fig. 4c-d), supporting Si-O and Si-ON bond formation on the surface of 6H-SiC, without Ga-oxide formation. Successive XPS measurements after graphene removal reveals no detectable chemical state change of the **N 1s**, **Ga 2p$_{3/2}$**, and **Ga 3d** core level spectra after 1 or 24 hour exposure to air (Supplementary Fig. 4e-h), while the peak intensities decrease due adsorption of adventitious carbon and contaminates from air exposure. The **N 1s** core level spectra indicate predominant Ga-N bonding from 2D GaN, with slight asymmetry on the high side of the binding energy due to Si-ON formation on the SiC substrate (Supplementary Fig. 4b). More importantly, no changes in the shape and energy position of the **Ga 2p$_{3/2}$** and **Ga 3d** core level spectra are observed with exposure to air (Supplementary Fig. 4f-h), supporting the chemical state stability of the 2D GaN in air, even without the overlying graphene capping layer.

## Supplementary References

## Figures

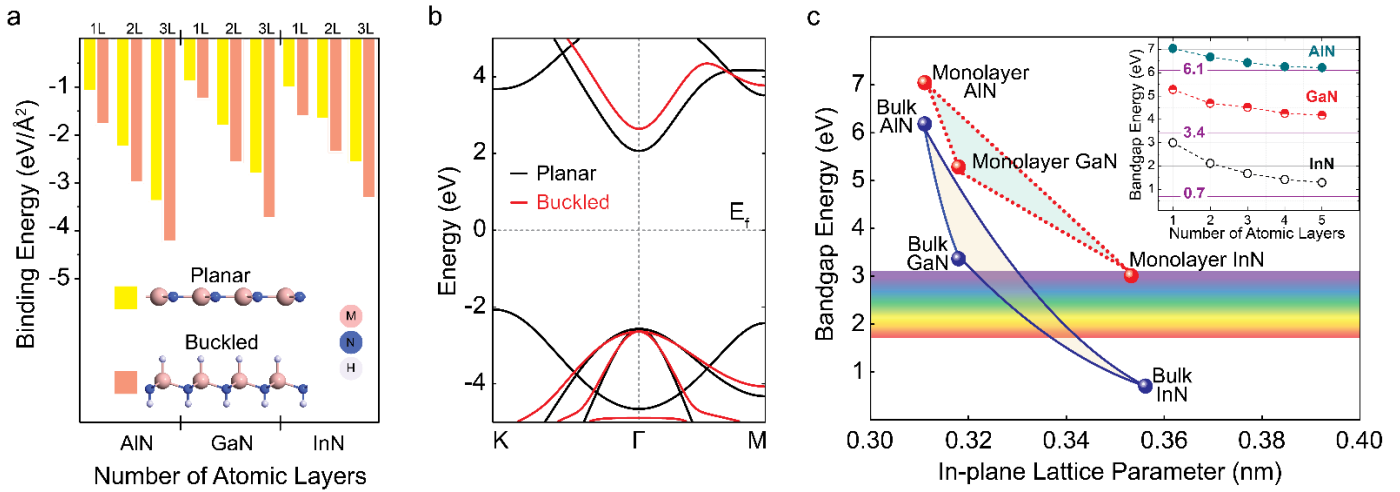

**Figure 1 | Properties of 2D nitrides from *ab initio* hybrid density functional theory. a**, binding energy calculations of freestanding planar and buckled 2D nitrides (M: group-III metal element, N: nitrogen and H: hydrogen atoms), as a function of layer number (L), showing increased stability of the buckled structure gleaned from their decreasing binding energies relative to planar 2D nitrides. **b**, bandstructure calculations *via* DFT *meta*-GGA of freestanding planar and buckled 2D monolayer GaN, illustrating buckled 2D GaN with a direct bandgap ($E_g$) of 5.28 eV and planar 2D GaN with indirect $E_g$ of 4.12 eV, both larger than the direct $E_g$ of wurtzite bulk GaN (3.42 eV) due to quantum confinement. **c**, diagram of bandgap energy *versus* in-plane lattice parameter for bulk and buckled 2D nitrides, establishing the possibility of probing deep into the ultraviolet with monolayers of group-III nitrides. The $E_g$ as a function of number of atomic buckled layers is included as an inset in (**c**).

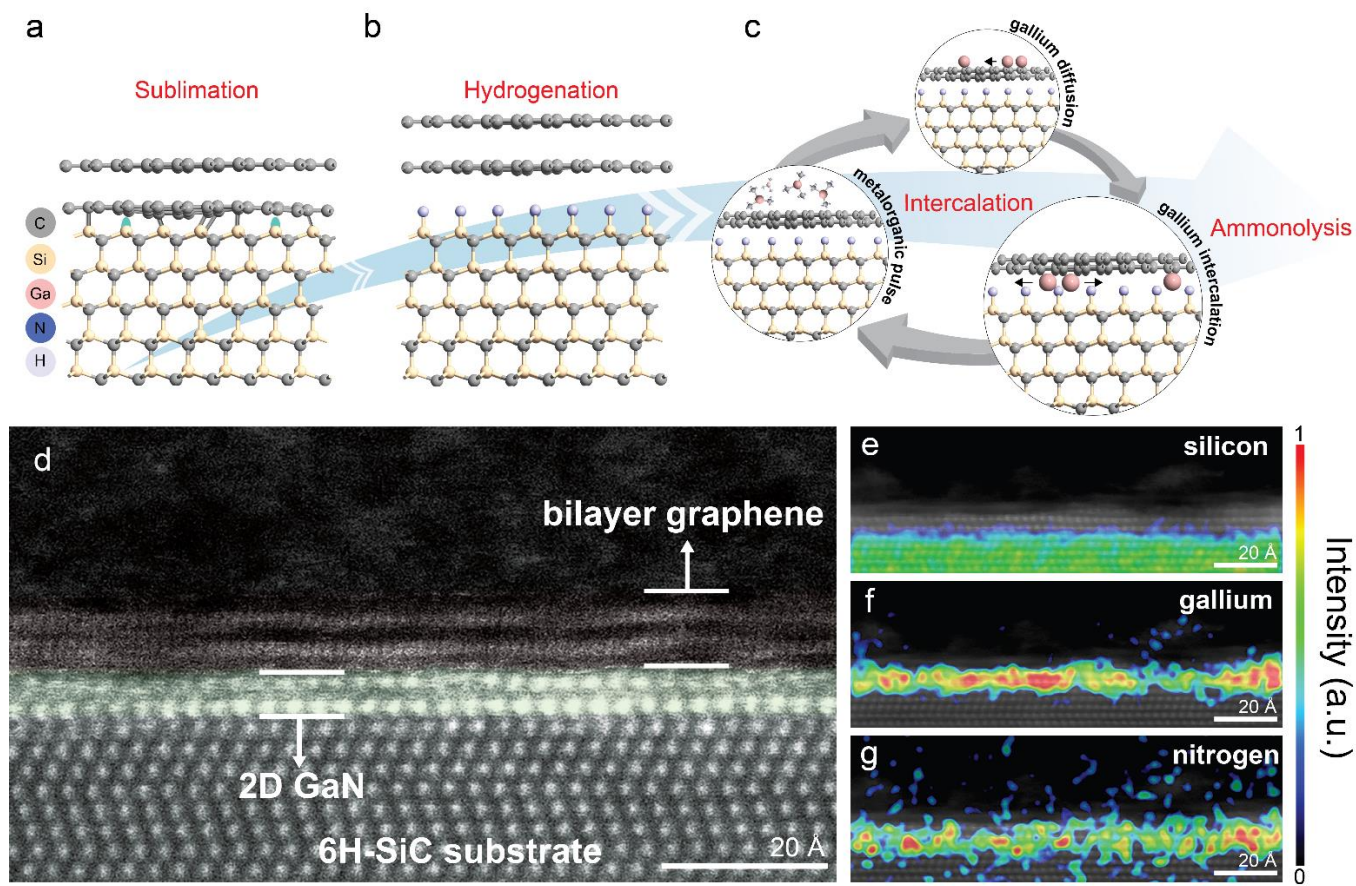

**Figure 2 | 2D GaN formation *via* migration-enhanced encapsulated growth (MEEG). a-c**, schematic of the MEEG process that leads to the formation of 2D GaN (**d**). **a**, the process of silicon sublimation from SiC(0001) to grow epitaxial graphene that consists of an initial partially bounded graphene-buffer layer (bottom) followed by a monolayer of graphene (top). **b**, exposing epitaxial graphene in (**a**) to ultrahigh purity hydrogen at elevated temperatures decouples the initial (bottom) graphene-buffer layer to form bilayer QFEG. **c**, the proposed MEEG process for the formation of 2D GaN: first, trimethylgallium precursor decomposition and gallium adatom surface diffusion; second, intercalation and lateral interface diffusion; finally, transformation of gallium to 2D GaN *via* ammonolysis. **d**, HAADF-STEM cross-section of 2D GaN, consisting of two sub-layers of gallium, between bilayer graphene and SiC(0001). **e-f**, elemental EDX mapping of silicon (**e**), gallium (**f**) and nitrogen (**g**) in 2D GaN.

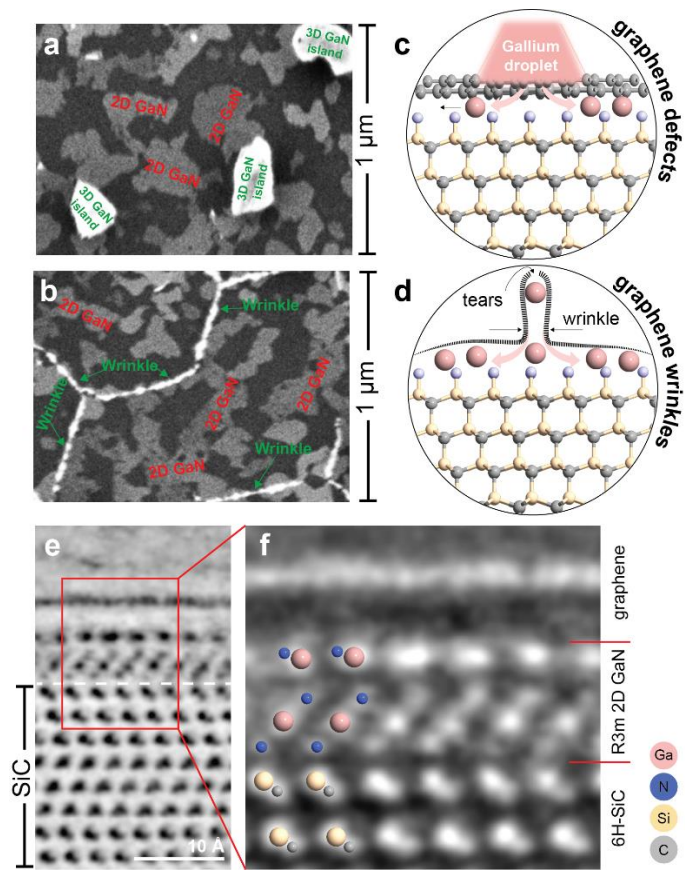

**Figure 3 | Pathways for intercalation and structure of 2D GaN. a-b**, SEM of 2D GaN (patchwork regions of bright contrast) concentrated near 3D GaN islands (**a**) and networks of graphene wrinkles (**b**). **c-d**, schematic of proposed pathways for intercalation through the accumulation and penetration of gallium through point defects in the graphene lattice (**c**) and through nanoscale tears at the apex of graphene wrinkles (**d**). **e-f**, annular bright field (ABF) images collected with aberration corrected STEM near the [11$\bar{2}$0] zone axis, resolving the atomic columns of gallium, nitrogen, silicon and carbon in graphene/2D GaN/6H-SiC(0001) heterostructure. **f**, is an inverted image of (**e**) for enhanced visualization. The R3m structure of 2D GaN is highlighted from the position of the nitrogen and gallium atomics columns.

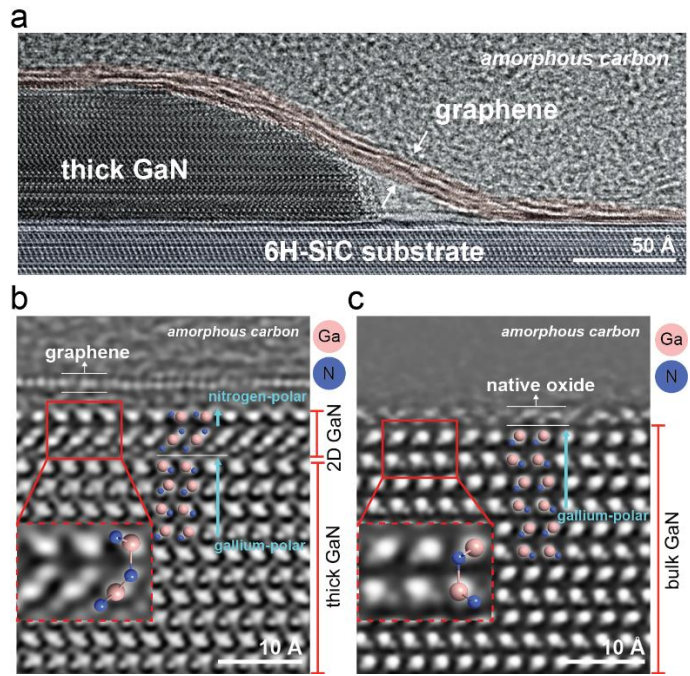

**Figure 4 | Role of graphene in the atomic stabilization of 2D nitrides. a**, TEM revealing the formation of thick GaN beneath graphene following the MEEG process and displaying graphene's notable ability to encapsulate GaN with little distortion to the inherit structure of graphene. **b-c**, inverted ABF-STEM images near the $[11\bar{2}0]$ zone axis, collected under identical conditions, comparing the interface of graphene and GaN (**b**) seen in (**a**) and the surface of bulk GaN grown directly on 6H-SiC(0001) without graphene capping (**c,** see Methods). **b**, the atomic columns of 2D GaN on thick GaN clearly shows the polarity inversion of the nitrogen-gallium termination in 2D GaN interfacing graphene (arrow labeled nitrogen-polar) from that of the polarity of thick GaN (arrow labeled gallium-polar). The role of graphene here in stabilizing the buckled R3m 2D structure and inverting polarity is undoubtedly evident from the displacement of (gallium and nitrogen) dumbells in the magnified insets. **c**, only reconstructed surface oxide is observed in bulk GaN grown directly on 6H-SiC(0001) without graphene capping and with no inherent changes in polarity (arrow labeled gallium-polar). In each inset (**c,d**), gallium-nitrogen positions and bonding (obtained from our DFT models) are overlaid on the micrographs.

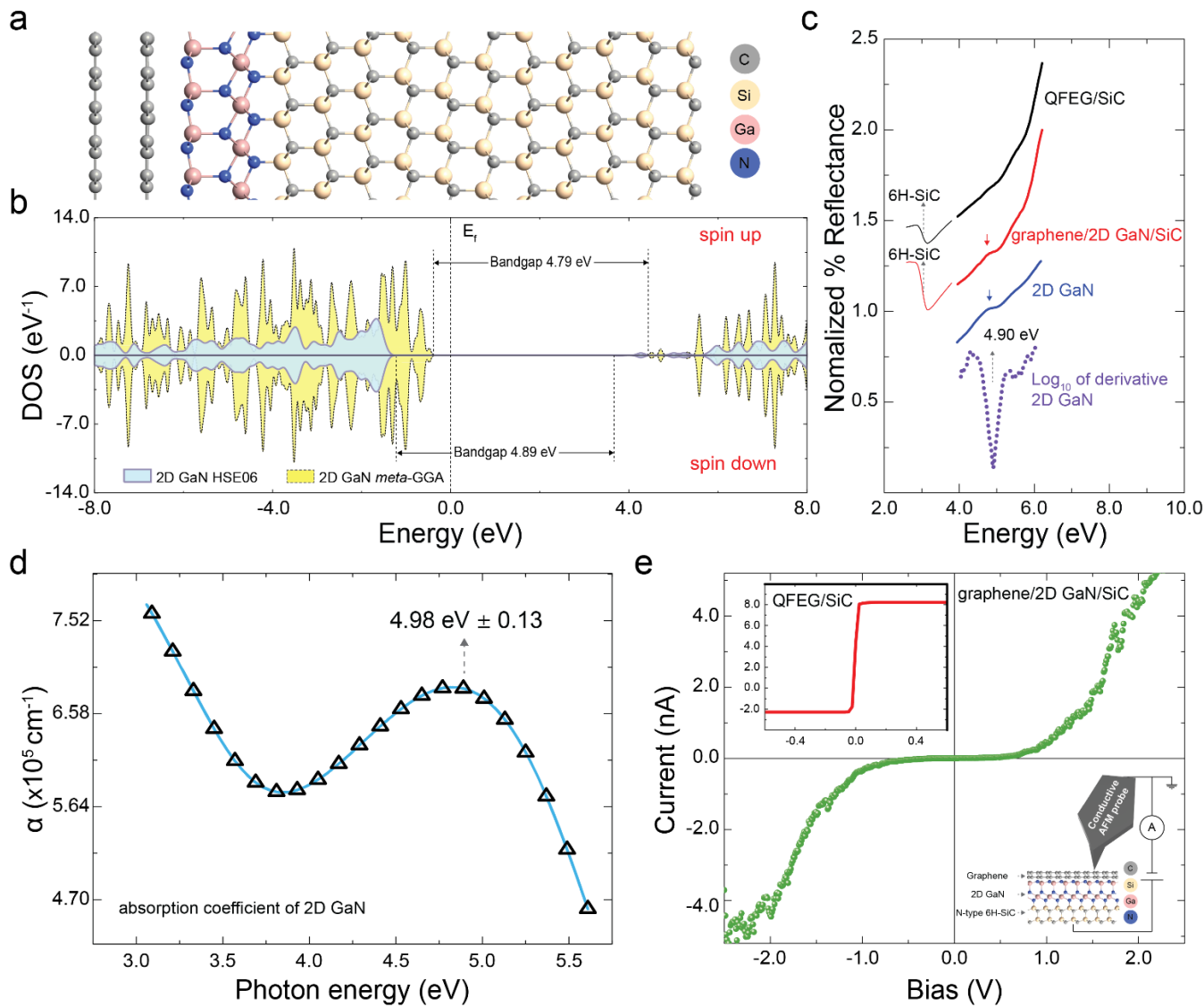

**Figure 5 | Density of state (DOS) calculations, bandgap ($E_g$) and electrical measurement of 2D GaN. a**, structure simulation with DFT of the graphene/2D GaN/6H-SiC heterostructure as observed in ABF-STEM (**Fig. 3e-f**). 2D GaN was modeled as a rhombohedral cell having R3m symmetry and optimized between bilayer graphene and 6H-SiC. **b**, DOS calculation *via* meta-GGA and HSE06 functionals, revealing the $E_g$ of 4.79 eV and 4.89 eV for 2D GaN from each approach, respectively. From the DOS in (**b**), the valance band shifts up towards the Fermi level ($E_f$), indicating p-type semiconducting behavior in 2D GaN **c**, UV-Vis reflectance collected with an integrating sphere. Spectra from top to bottom: QFEG/SiC; graphene/2D GaN/SiC; 2D GaN after QFEG/SiC background subtraction and the derivative of the 2D GaN extracted reflectance plotted in $Log_{10}$ scale, revealing the $E_g$ of 2D GaN is 4.90 eV. **d**, absorption coefficient (α) of 2D GaN, collected with UV-Vis spectroscopic ellipsometry, revealing a direct $E_g$ of 4.98 eV. **e**, Vertical transport measurements with 2D GaN Schottky barrier. I-V curve of the heterostructure (green curve) and QFEG/6H-SiC (red inset curve) collected with conductive AFM.

**Supplementary Figures**

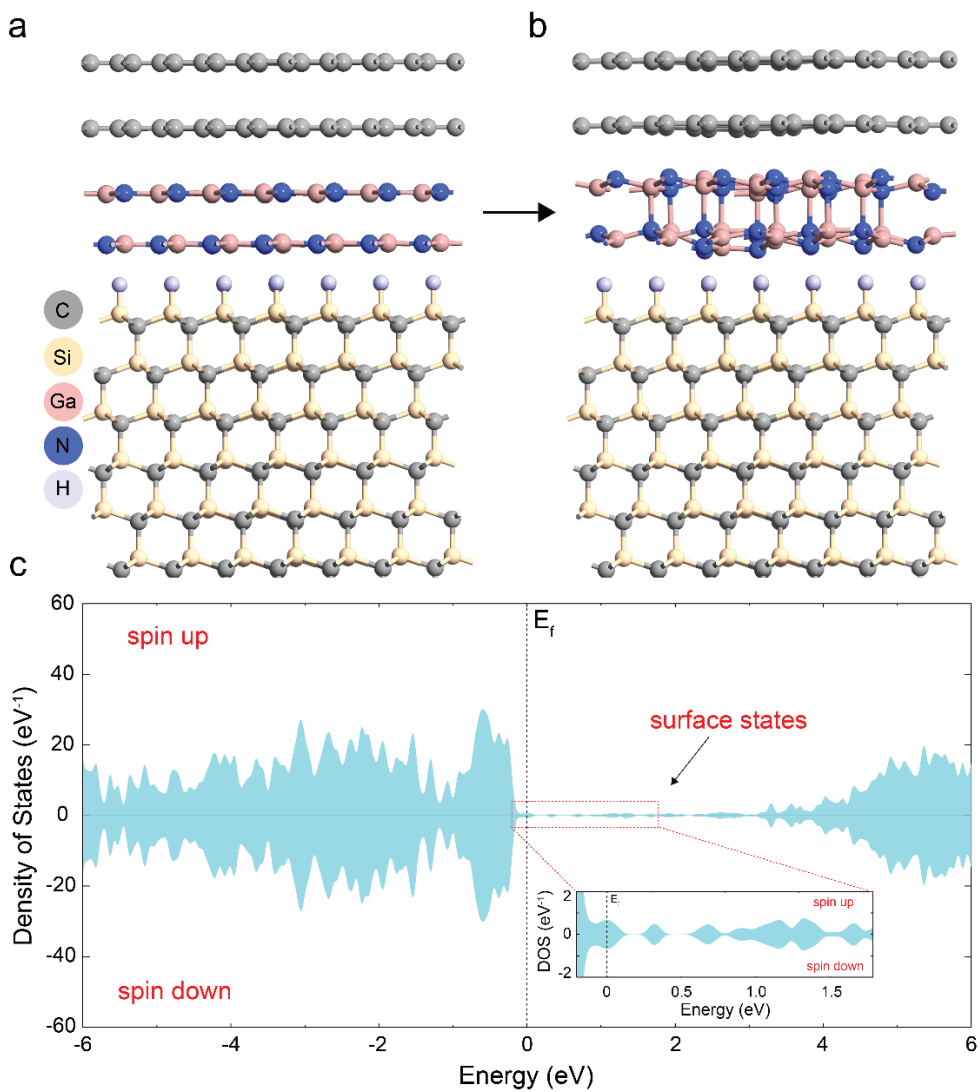

**Supplementary Figure 1 | Planar 2D bilayer nitrides lead to surface states. a-b**, structure simulation of planar 2D bilayer GaN before (**a**) and after (**b**) structure relaxation within a bilayer graphene/6H-SiC(0001) heterostructure. **b-c**, after optimization, clear bonding between the initial van der Waals (vdW) separated planar layers is evident and clearly reflected in its DOS (**c**). **c**, DOS calculations after relaxation of planar 2D bilayer GaN in the supercell (**b**), revealing surface states near the Fermi level ($E_f$) and therefore, semi-metallic behavior of the 2D GaN structure that prefers buckling over remaining planar.

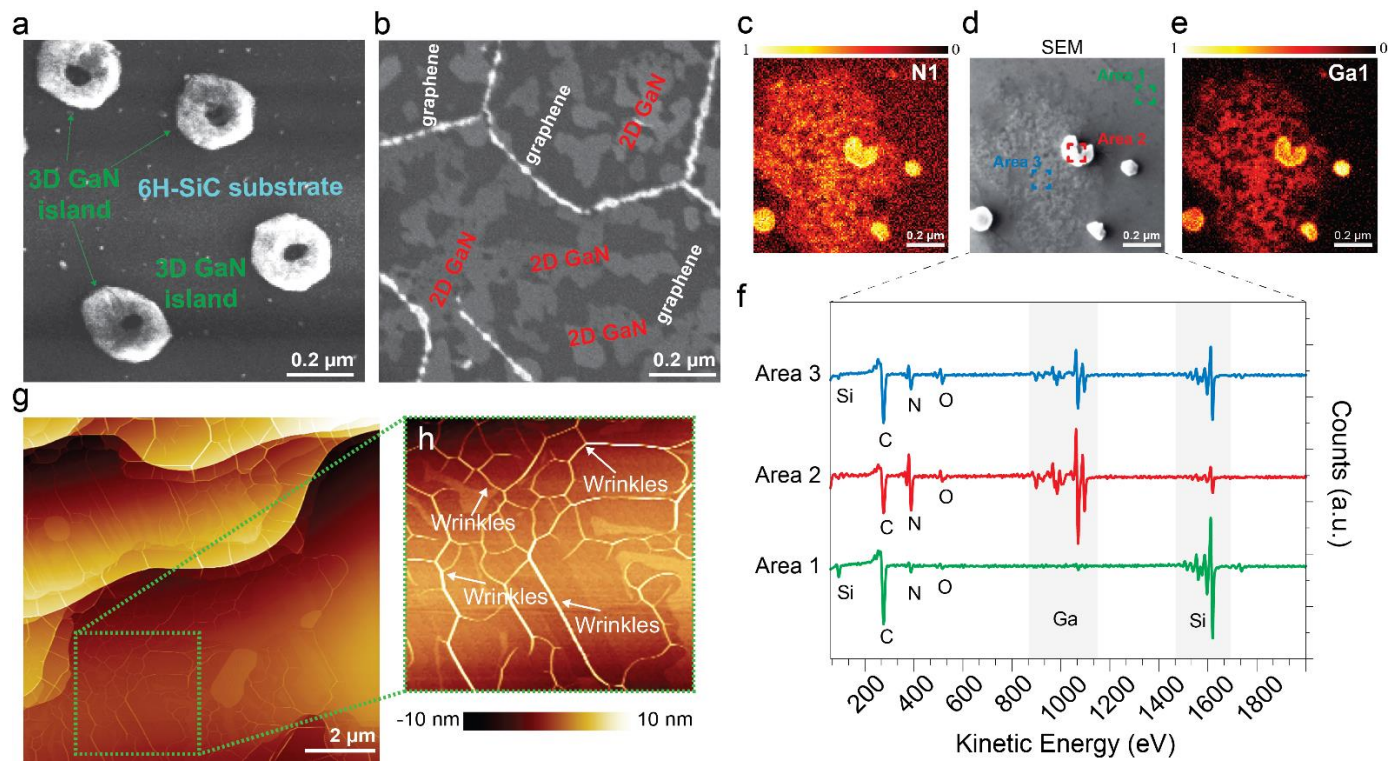

**Supplementary Figure 2 | Characterization of 2D GaN and graphene. a**, SEM showing the formation of 3D doughnut-shaped islands of GaN on SiC(0001) *via* MEEG when graphene capping is not utilized. **b**, SEM showing the formation 2D GaN on SiC *via* MEEG when graphene capping (QFEG) is utilized. Patchwork of regions of bright contrast (**b**) are of 2D GaN. **c-f**, regions of bright contrast in SEM (**d**) identified as area of nitrogen (**c**) and gallium (**e**) from surface sensitive elemental maps collected in high resolution scanning Auger electron spectroscopy (AES). **f**, AES survey spectra of kinetic energy *versus* counts of three selected areas in SEM (**d**) labeled areas 1, 2 and 3. From the SEM contrast, Area 1 is a region of bare QFEG on SiC outside the regions of 2D GaN, as clearly reflected in the AES survey spectrum by the presence of negligible gallium and nitrogen signal when compared to Areas 2 and 3. In the case of Areas 2 (3D GaN island on the surface of graphene) and Area 3 (region of 2D GaN) in SEM (**d**), their corresponding AES spectra are clearly representative of the relative changes of gallium and nitrogen counts collected from those regions. **g-f**, AFM of the surface of epitaxial graphene, illustrating the dense network of graphene wrinkles.

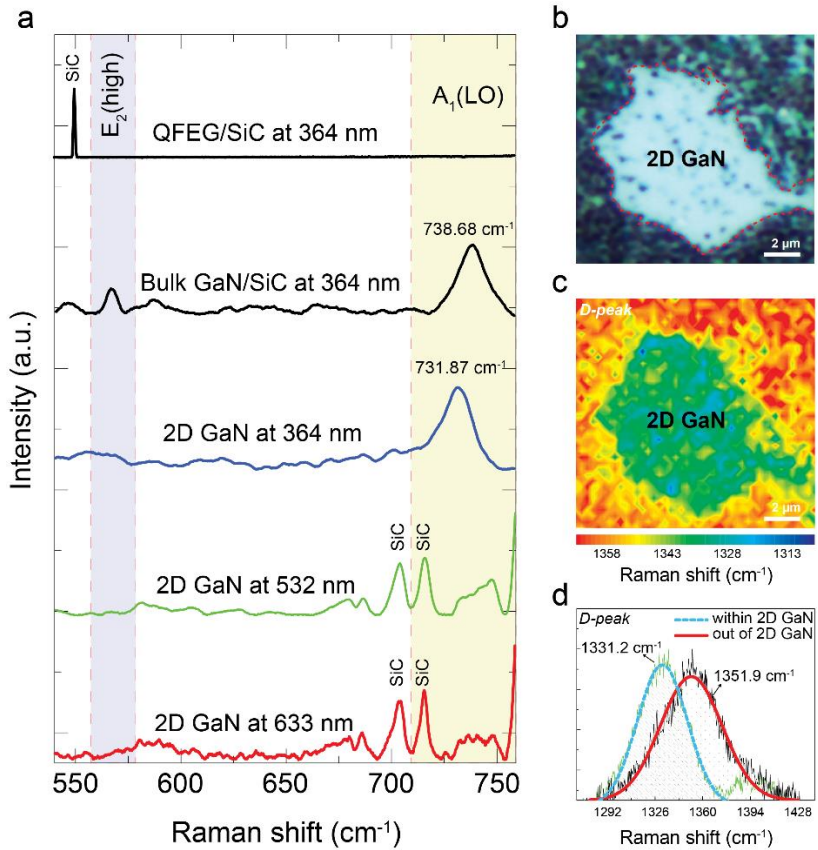

**Supplementary Figure 3 | Raman characterization of 2D GaN and graphene. a**, comparison of Raman spectra (from top to bottom): QFEG/6H-SiC and bulk GaN/6H-SiC (black curves collected at 364 nm laser excitation), to the heterostructure consisting of graphene/2D GaN/6H-SiC collected at varied laser excitation (blue, green and red curves at 364 nm, 532 nm and 633 nm, respectively). The shallow penetration depth of UV Raman (364 nm laser excitation) enhances surface sensitivity to 2D GaN and improves Raman scattering efficiency while effectively suppressing active Raman phonon modes from 6H-SiC and graphene (Supplementary Reference 10). In 2D GaN, a longitudinal optical phonon mode $A_1(LO)$ appears at 731.87 cm$^{-1}$, blue-shifted from the $A_1(LO)$ mode (738.68 cm$^{-1}$) for bulk GaN at the same laser excitation energy and power. The $E_2$(high) mode in bulk GaN with 364 nm laser excitation does not appear in 2D GaN, regardless of the laser excitation. **b**, optical micrograph of the sample region containing 2D GaN within the heterostructure and **c**, Raman map (**c**) of the position of the graphene defect (D)-peak (at 532 nm laser excitation energy), illustrating **e**, a large blue-shift in the Raman D-peak position relative to non-intercalated regions of graphene (~1351.9 cm$^{-1}$ to ~1331.2 cm$^{-1}$). This is largely due to structural changes in the graphene and potential interactions at the graphene/2D GaN interface (Supplementary Reference 11).

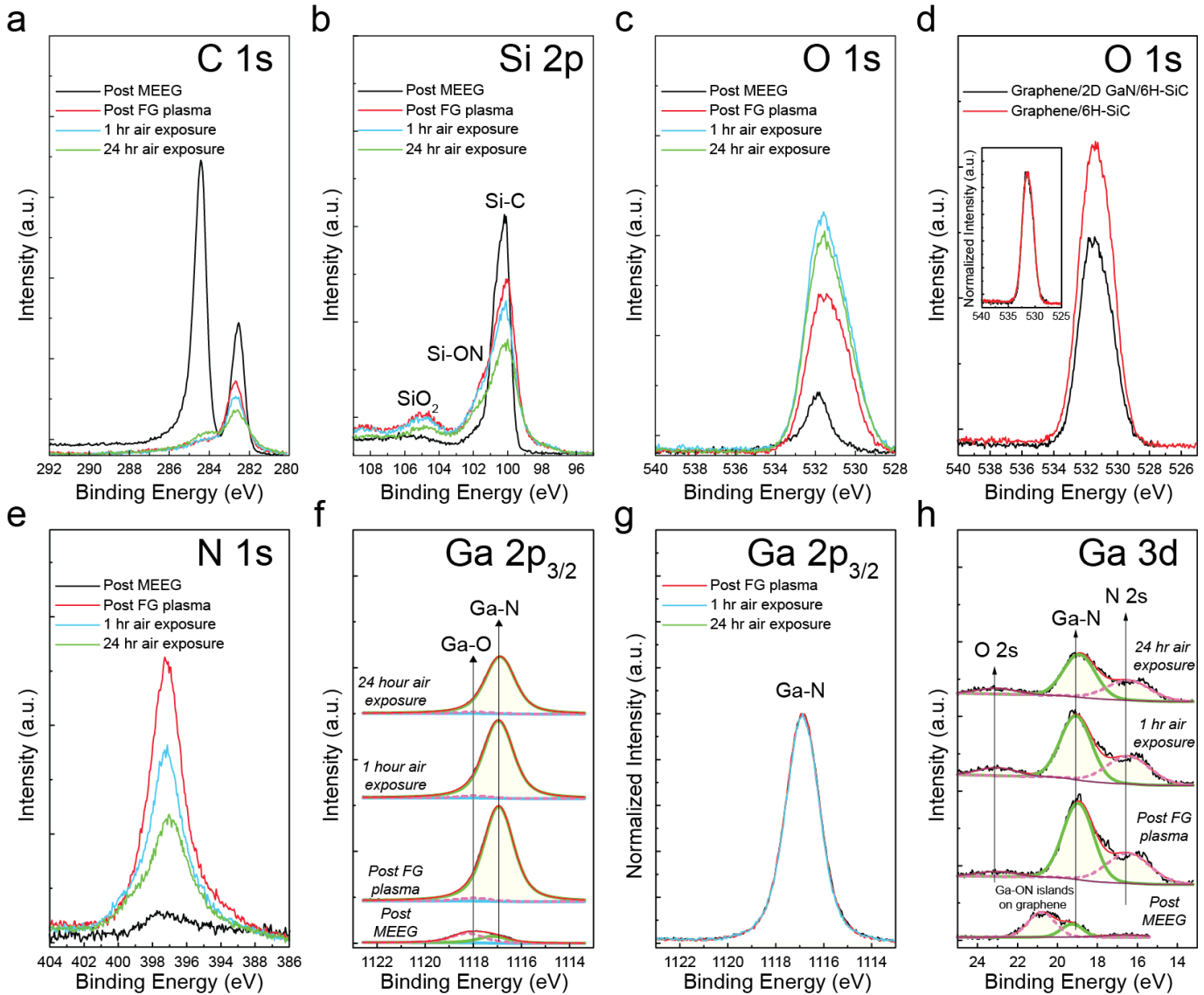

**Supplementary Figure 4 | Stability of 2D GaN monitored by x-ray photoelectron spectroscopy (XPS).** High-resolution XPS spectra collected after the growth of 2D GaN (Post MEEG, black); after removal of graphene with forming gas plasma (Post FG plasma, red); after exposure to air for 1 hour (blue) and 24 hours (green) for the core-levels of **a**, C 1s. **b**, Si 2p. **c-d**, O 1s, also comparing the intensity and shape (normalized inset) of the QFEG/6H-SiC (red) and graphene/2D GaN/6H-SiC (black) after exposure to FG plasma (**d**). **e**, N 1s. **f-g**, Ga 2p3/2, where (**g**) shows no changes in the Ga-N bonding after exposure to air. **h**, shows the Ga 3d core-level spectra. From deconvolution of the Ga $2p_{3/2}$ (**f**) and Ga 3d (**h**) core-level spectra, no detectible changes in the chemical state is observed. Therefore, 2D GaN is stable in air after removal of graphene from the surface of the graphene/2D GaN/6H-SiC heterostructure in FG plasma (see Methods and Supplementary Note c).

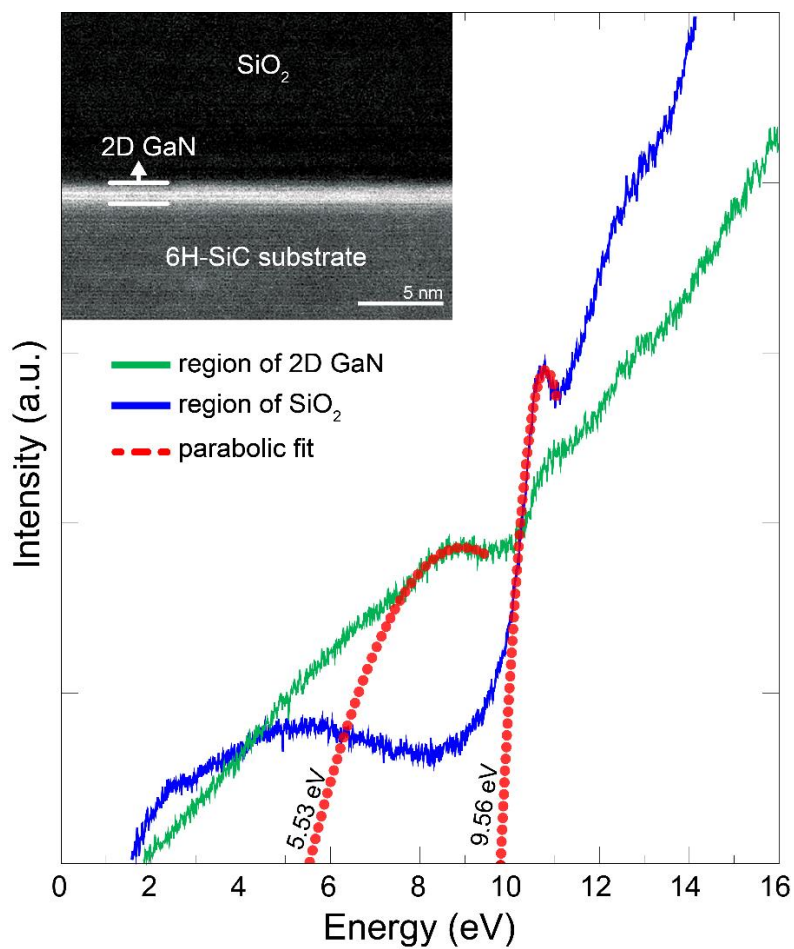

**Supplementary Figure 5 | Bandgap ($E_g$) measurements *via* low-loss electron-energy loss spectroscopy (EELS).** Point spectrum of low-loss EELS collected from regions of interest in the STEM image shown in the inset. The $E_g$ is extracted from the low-loss EELS spectrum after background subtraction using the parabolic fitting method (red curves) of the first prominent rise. The blue curve is spectrum collected in the bulk of the $SiO_2$ protective layer and the green curve is spectrum collected in the region of 2D GaN, which show an extracted $E_g$ of 9.56 eV and 5.53 eV, respectively (see Methods and Supplementary Note b).